\documentclass[twocolumn,showpacs,aps,prl,superscriptaddress]{revtex4b5}

\usepackage{graphicx}
\usepackage{dcolumn}
\usepackage{amsmath}
\usepackage{epsfig}

\input pubboard/babarsym
\def\Dp      {\ensuremath{D^+}}
\def\Dm      {\ensuremath{D^-}}
\def\Dstar   {\ensuremath{D^{*}}}
\def\Dstarp  {\ensuremath{D^{*+}}}
\def\Dstarm  {\ensuremath{D^{*-}}}

\def\rhop    {\ensuremath{\rho^+}}
\def\rhom    {\ensuremath{\rho^-}}
\def\aonep   {\ensuremath{a_1^+}}

\newcommand{\BABARPubYear}    {01}
\newcommand{\BABARPubNumber}  {06}

\newcommand{\SLACPubNumber} {8847}

\def\figurebox#1#2#3{%
    \def\arg{#3}%
    \ifx\arg\empty
    {\hfill\vbox{\hsize#2\hrule\hbox to #2{\vrule\hfill\vbox to #1{\hsize#2\vfill}\vrule}\hrule}\hfill}%
    \else
    {\hfill\epsfbox{#3}\hfill}%
    \fi}

\newcommand{\deltaE} {\ensuremath{\Delta \rm{E}}\xspace}
\newcommand{\Brec} {\ensuremath{\B_{\rm{rec}}}\xspace}
\newcommand{\Bopp} {\ensuremath{\B_{\rm{opp}}}\xspace}

\def\Dstarp   {\ensuremath{D^{*+}}\xspace}
\def\Dstarm   {\ensuremath{D^{*-}}\xspace}

\def\DorDstarb  {\ensuremath{\Dbar^{(*)}}\xspace}

\def\DorDstarm {\ensuremath{D^{(*)-}}\xspace}
\def\DorDstarzb  {\ensuremath{\Dbar^{(*)0}}\xspace}

\begin{document}

\begin{flushleft}
\babar-PUB-\BABARPubYear/\BABARPubNumber\\
SLAC-PUB-\SLACPubNumber\\
\end{flushleft} 

\title{
\vskip 8mm
{\large \bf Measurement of the \Bz and \Bu meson lifetimes
with fully reconstructed hadronic final states}
\vskip 10mm
\begin{center}
{The \babar\ Collaboration}
\end{center}
}

%
\author{B.~Aubert}
\author{D.~Boutigny}
\author{J.-M.~Gaillard}
\author{A.~Hicheur}
\author{Y.~Karyotakis}
\author{J.~P.~Lees}
\author{P.~Robbe}
\author{V.~Tisserand}
\affiliation{Laboratoire de Physique des Particules, F-74941 Annecy-le-Vieux, France }
\author{A.~Palano}
\affiliation{Universit\`a di Bari, Dipartimento di Fisica and INFN, I-70126 Bari, Italy }
\author{G.~P.~Chen}
\author{J.~C.~Chen}
\author{N.~D.~Qi}
\author{G.~Rong}
\author{P.~Wang}
\author{Y.~S.~Zhu}
\affiliation{Institute of High Energy Physics, Beijing 100039, China }
% ...
\collaboration{The \babar\ Collaboration}
\noaffiliation

\date{July 9, 2001}

\begin{abstract}
The \Bz\ and \Bu\ meson lifetimes have been measured in \epem\ annihilation 
data collected in~1999 and 2000 with the \babar\ detector at center-of-mass energies
near the \FourS resonance. Events are selected in which one \B\ meson is
fully reconstructed in a hadronic final state while the second 
\B~meson is reconstructed inclusively. A combined fit to the \Bz\ and 
the \Bu\ decay time difference distributions yields
$ \tau_{\Bz} = 1.546 \pm 0.032\mbox{ (stat)}\pm 0.022\mbox{ (syst)}\ps$, 
$ \tau_{\Bu} = 1.673 \pm 0.032\mbox{ (stat)}\pm 0.023\mbox{ (syst)}\ps$
and $ \tau_{\Bu} / \tau_{\Bz} = 1.082 \pm 0.026\mbox{ (stat)}\pm 0.012\mbox{ (syst)}$.
\end{abstract}

\pacs{13.25.Hw, 12.39.Hg}

\maketitle

The spectator quark model predicts that the two charge states of a
meson with one heavy quark~$Q$ ($Q\ubar$~and~$Q\dbar$) have the same lifetime.
Deviations from this simple picture are expected to be
proportional to~$1/m_Q^2$~\cite{bib:Bigi, bib:Neubert}. Therefore, 
any lifetime differences are anticipated to be
much smaller for bottom than for charm mesons.
Various models~\cite{bib:Bigi, bib:Neubert} predict the ratio
of the \Bu and \Bz meson~\cite{bib:conjugate} lifetimes to differ by
up to 10\% from unity.
At present, this ratio is measured to be
$\tau_{\Bu}/\tau_{\Bz} = 1.062\pm 0.029$~\cite{bib:PDG2000}, with 
the most precise
values obtained by experiments operating near the~$Z$ and at
hadron colliders.

The lifetime measurements described here are 
based on a sample of approximately 23~million \BB pairs recorded near the \FourS resonance with
the \babar\ detector at the Stanford Linear Accelerator Center. 
The \pep2\ asymmetric-energy \epem\ collider produces $B^+ B^-$ and 
\BzBzb\ pairs moving along the beam axis ({\it
z} direction) with a nominal Lorentz boost of $\beta \gamma = 0.56$. Hence, on
average, the two \B\ decay vertices are separated by 
$\langle |\deltaz|\rangle = \beta \gamma \gamma_B^{\rm cms} c \tau \approx 270 \mum$, 
where~$\tau$ is either the \Bz~or \Bu~lifetime, and~$\gamma_B^{\rm cms}$
is the Lorentz factor of the \B~mesons in the \FourS rest frame.
This separation allows \B~lifetimes to be measured at the~\FourS, with
good statistical precision and systematic error sources
different from those in previously published results.

In this analysis, one of the \B mesons in an event, denoted 
\Brec, is fully reconstructed in a variety of
two-body charm and charmonium final
states. The decay point of the other 
\B\ in the event, 
\Bopp, is reconstructed
inclusively.  The decay probability distribution is given by
\begin{equation}
\label{eq:decayproba}
g(\deltat  | \tau) = \frac{1}{N} \cdot \frac{{\rm d}N}{{\rm d}(\deltat )} = 
\frac{1}{2 \tau} \  e^{-|\deltat |/\tau},
\end{equation}
where $\deltat  = t_{\rm rec}-t_{\rm opp}$ is the (signed) difference 
of the proper decay times of the \B\ mesons. The time interval~\deltat 
between the two \B decays is determined from~\deltaz, including an 
event-by-event correction for the direction of the \B~mesons with respect to 
the $z$~direction in the \FourS frame.
The challenge of the measurement is to
disentangle the resolution in \deltaz, 190\mum on average,
from the effects of the \B lifetime, since
both contribute to the width of the \deltat~distribution. 
In the absence of background, the measured \deltat\ distribution
is described by the probability density function (PDF)
\begin{equation}
\label{eq:Phi}
{\cal G}(\deltat,\sigma | \tau,\hat{a}) = \int_{-\infty}^{+\infty}g(\deltat'
| \tau) \, \calR \it (\deltat-\deltat', \sigma |\hat{a}) \; {\rm d}(\deltat'),
\end{equation}
where $\cal R$ is the \deltat\ resolution function with
parameters~$\hat{a}$, and $\sigma$ is the event-by-event error
on~$\deltat$ calculated from the vertex fits.
An unbinned maximum likelihood fit is used to extract the \Bz and \Bu
lifetimes from the \deltat~distributions for \BzBzb\ and $B^+ B^-$
events.

The \babar\ detector is described in detail elsewhere~\cite{bib:babar}. 
Charged particle trajectories are measured by a combination of a silicon vertex 
tracker~(SVT) and a drift chamber~(DCH) in a 1.5-T solenoidal field. 
For 1\gevc~tracks, the impact parameter resolutions in~$z$ and in
the transverse plane are~65\mum\ and~55\mum, respectively.
Photons and electrons are detected in
the CsI(Tl) electromagnetic calorimeter (EMC). A ring imaging Cherenkov
detector, the DIRC, is used for charged hadron identification. The DCH and SVT 
also provide ionization measurements, \dedx, for particle identification. 
The instrumented flux return (IFR) is segmented and contains resistive plate
chambers to identify muons. 
Electron candidates are required to have a ratio of EMC
energy to track momentum, 
an EMC cluster shape, DCH \dedx\ and DIRC Cherenkov angle
consistent with the electron hypothesis.
Muon candidates are required to have an energy deposit in the EMC 
consistent with the muon hypothesis, IFR~hits located consistently 
on the extrapolated DCH~track, and an IFR penetration in interaction
lengths consistent with the muon hypothesis.

\Bz\ and \Bu\ mesons are reconstructed in a sample of multihadron events in the
modes $\Bz \ra \DorDstarm \pip$, $\DorDstarm \rho^+$, $\DorDstarm a_1^+$,
$\jpsi  \Kstarz$ and $\Bu \ra \DorDstarzb \pip$, $\jpsi K^+$, $\psitwos K^+ $. 
Multihadron events must have
a minimum of three reconstructed charged tracks, 
a total charged and neutral energy greater than
4.5\gev, and an event vertex 
within 0.5\cm\ of the beam spot~\cite{bib:babar} center 
in~$xy$ and within 6\cm\ in~$z$. The event vertex is determined
from all charged tracks that have an impact parameter 
with respect to the beam spot center smaller than 1\cm 
in~$xy$ and 3\cm in~$z$.

For \piz\ candidates, pairs of photons in the EMC, each with more than 
30\mev of energy, are selected if their invariant mass is  
within 20\mevcc\ of the \piz\ mass~\cite{bib:PDG2000} and their total 
energy exceeds 200\mev\ (100\mev\ for the soft \piz\ in \Dstar\ decays).
A mass constraint is applied to selected candidates for use in the 
subsequent reconstruction chain.

$\KS\to\pip\pim$ candidates are required to have an
invariant mass between 462 and 534\mevcc. 
A geometrical vertex fit with $\chi^2$~probability
above~0.1\% is required, and the transverse flight distance
from the event vertex must be greater than 0.2\cm.

\Dzb\ candidates are reconstructed in the 
decay channels $\Kp\pim$, $\Kp\pim\piz$, $\Kp\pip\pim\pim$ and $\KS
\pip\pim$ and \Dm\ candidates in the decay channels
$\Kp \pim \pim$ and $\KS\pim$. 
Kaons from \Dm~decays and charged daughters from $\Dzb\to\Kp \pim$ 
are required to have momenta greater than 200\mevc. All other charged
\Db~daughters are required to have momenta greater than 150\mevc.
For $\Dzb\to\Kp \pim \piz$, we only reconstruct the dominant resonant
mode, $\Dzb\to\Kp \rhom$, followed by $\rhom\to\pim \piz$. The 
$\pim \piz$ mass is required to lie within 150\mevcc\ of the 
$\rho$~mass~\cite{bib:PDG2000} and
the angle between the \pim\ and \Dzb\ in the $\rho$ rest frame, 
$\theta^*_{\Dz\pi}$, must satisfy $|\cos \theta^*_{\Dz \pi}| > 0.4$.
All \Dzb\ and \Dm\ candidates are required to have a momentum greater
than 1.3\gevc\ in the \FourS frame, 
an invariant mass within $3 \sigma$ of the 
nominal value~\cite{bib:PDG2000} and a geometrical vertex fit with 
a $\chi^2$~probability greater than~0.1\%. A~mass constraint is
applied to selected \Db candidates.

Charged and neutral \Dstarb\ candidates are formed by combining 
a \Dzb\ with a \pim\ or \piz. The momentum of the pion in the \FourS
frame is required to be less than 450\mevc.
The soft \pim\ is constrained to originate from the beam spot
when the \Dstarm\ vertex is fit. 
After the mass constraint to the \Dzb\ daughter, \Dstarb\ candidates
with $m(\Dzb\pi)$ within 2.5$\sigma$ of the nominal
mass~\cite{bib:PDG2000} for \Dstarm, or within 4$\sigma$ of the nominal
mass~\cite{bib:PDG2000} for \Dstarzb, are selected.

Candidates for leptonic decays of charmonium mesons must have at least one 
decay product positively identified as an electron or a muon.
If it traverses the calorimeter, the second muon must be consistent
with a minimum ionizing particle.
$\jpsi$ candidates are required to lie in the invariant mass interval 
2.95 (3.06) to 3.14\gevcc\ for the $\epem$ ($\mu^+\mu^-$) channel.
The $\epem$ ($\mu^+\mu^-$) invariant mass of \psitwos\ candidates
must be between 3.44 (3.64) and 3.74\gevcc. A mass constraint is applied 
to selected candidates. $\psitwos \ra \jpsi \pip \pim$ candidates 
are selected if the $\pip \pim$~mass is between
0.4 and 0.6\gevcc and the \psitwos mass is
within 15\mevcc\ of the nominal value~\cite{bib:PDG2000}. All
\psitwos\ candidates must have momenta between 1.0 and 1.6\gevc\
in the \FourS\ rest frame.

\B\ candidates are formed by combining a \DorDstarb, \jpsi\ or
\psitwos\ candidate
with a \pip, \rhop, \aonep\ $(\aonep\to\pip\pim\pip)$, 
\Kstarz\ $(\Kstarz\to\Kp\pim)$ or $K^+$ candidate that has a momentum
larger than 500\mevc\ in the \FourS frame.
For $\Bz \to \DorDstarm \rhop$, the \piz\ from the \rhop\ decay
must have an energy greater than 300\mev. For 
$\Bz\to\DorDstarm\aonep$, the \aonep\ must have an 
invariant mass between
1.0 and 1.6\gevcc, and the $\chi^2$~probability of a vertex fit of the
\aonep\ candidate is required to be greater than~0.1\%. 
Positive identification of kaons is required for modes with higher
background, such as $\Bu\to\Dstarzb\pip$ with $\Dzb\to\Kp\pip\pim\pim$.

Continuum $e^+e^-\to q\overline q$ background is rejected by requiring
the normalized second Fox-Wolfram moment~\cite{bib:FoxW} for the event
to be less than 0.5.
Further suppression is achieved by a mode-dependent
restriction on the angle between the \Brec\ and \Bopp
thrust axes in the \FourS~frame.

\Bz\ and \Bu\ candidates are identified on the basis of 
the difference \deltaE between the reconstructed energy 
and the beam energy~$\sqrt{s}/2$ in the \FourS\ frame, and the
beam-energy substituted mass
\mes\ calculated from $\sqrt{s}/2$ and the reconstructed momentum 
of the candidate. \B~candidates are selected with $\mes > 5.2$\gevcc
and $|\deltaE| < 3 \sigma_{\deltaE}$, where
$\sigma_{\deltaE}$ (10 to 30 MeV) is the measured 
resolution for each decay mode.

The decay position of the \Brec\ candidate is determined by requiring 
convergence of a vertex fit, where in addition the masses of the 
$D$ mesons are constrained to their nominal values~\cite{bib:PDG2000}. 
Precisions between 60 and 100\mum\ {\it rms} for the
\Brec decay position in~$z$ and in the transverse plane are achieved,
depending on the decay mode.

\begin{figure}
\includegraphics[width=0.90\linewidth,height=0.8365\linewidth]{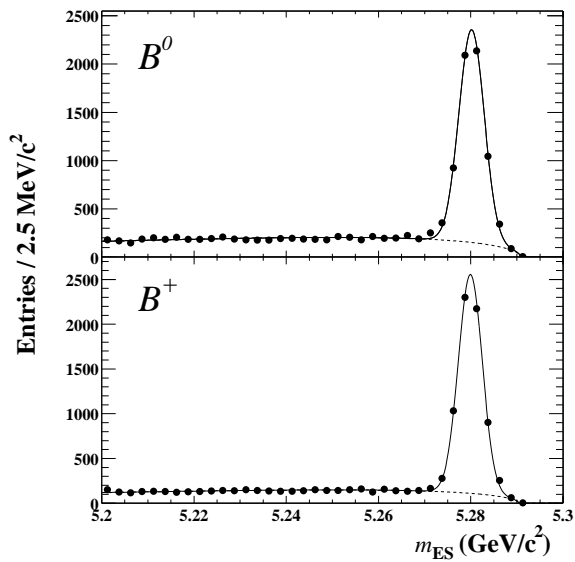}
\caption{\label{fig:breco}
\mes\ distributions of the selected neutral (top) and charged (bottom)
\Brec candidates.}
\end{figure}

The vertex of the \Bopp\ is determined from all tracks in the event
after removing those 
associated with the \Brec~candidate. Tracks from
photon conversion candidates are rejected. Daughter tracks from
\KS\ or $\Lambda$ candidates are replaced by the neutral parents.
An additional constraint is imposed on the \Bopp\ vertex using the \Brec\ vertex and
three-momentum, the beam spot position, and the average \FourS momentum.
To reduce the bias in the forward $z$~direction from charm decay
tracks, the track with the largest contribution to the vertex $\chi^2$, if
above~6, is removed and the fit iterated until
no track fails this requirement.
Events are required to have at least 2 tracks remaining
in the \Bopp\ vertex, an error on $\deltaz$ 
smaller than 400\mum and $|\deltaz |< 3000\mum$.
The precision achieved on \deltaz, 190\mum\ {\it rms} on average, is
dominated by the resolution on the \Bopp\ vertex.
A~remaining bias of $-35\mum$ due to charm decays on the \Bopp\ side is
observed. 
We require~$|\deltat |< 18\ps$ and find $6064 \pm\ 70$ \Bz\ and $6336 \pm\ 63$
\Bu\ signal events in a $\pm 2.5 \sigma$ ($\sigma=2.7\mevcc$ and
$2.6\mevcc$, respectively) window around
the \mes~peak above a small background ($\simeq 10\%$).
The \mes\ distributions for the final samples are shown in
Fig.~\ref{fig:breco} along with the results of a fit with 
a Gaussian distribution for the signal and an ARGUS background
function~\cite{bib:argusfunction}. 

As already noted, the modeling of the resolution
function~$\cal{R}$ is a crucial element of the \B\ lifetime
measurements. Studies with both Monte Carlo simulation and data show
that the sum of a zero-mean Gaussian distribution and its convolution 
with a decay exponential 
provides a good trade-off between different sources of 
uncertainties: \par 

\small{
\vspace*{-6mm}
\begin{eqnarray}
\label{eq:resol}
{\cal R} (\delta_{t} , \sigma | \hat{a}= \{ h,s,\kappa \}) = 
h    \frac{1}{\sqrt{2\pi}  s \sigma}   \exp\left(
-\frac{\delta_{t}^2}{2s^2 \sigma^2}
\right) \ \ \ \ \ \ \ \ \ \ \ \ \ \\
 + \int_{-\infty}^{0} \frac{1-h}{\kappa\sigma}  \exp\left(
\frac{\delta_{t}'}{\kappa\sigma} \right)   
\frac{1}{\sqrt{2\pi}  s \sigma}   \exp\left(
-\frac{(\delta_{t} - \delta_{t}')^2}{2s^2 \sigma^2}\right)
{\rm d}(\delta_{t}')\; ,\nonumber 
\end{eqnarray}
}
\normalsize
\hspace*{-5pt}
where $\delta_{\rm t}$ is the difference between the
measured and true $\deltat$ values. 
The model parameters $\hat{a}$ are  
the fraction $h$ in the core Gaussian component, 
a scale factor $s$ for the per-event errors $\sigma$, and the 
factor~$\kappa$ in the effective
time constant~$\kappa\sigma$ of the exponential that accounts for the
effect of charm decays.
Monte Carlo studies show that the parameters $\hat{a}$ obtained for 
different decay modes are compatible, as expected for
\deltat~resolution dominated by the \Bopp\ vertex.
The resolution functions for \Bz and \Bu mesons
differ slightly because the \Bopp decays to a different admixture of $D^-$ and
\Dzb\ mesons. The difference is not significant given the present data
sample size. Hence a single set of resolution function parameters is
used for both \Bz\ and~\Bu\ in the lifetime fits, and a small
correction discussed later is applied to the results.
While the resolution function ${\cal R}$ describes almost all events,
incorrectly measured {\it outlier} events are modeled separately as
discussed below.

The unbinned maximum likelihood fit for the \B~lifetimes uses
all events with $\mes > 5.2\gevcc$.
The probability $p_i^{sig}$ for event $i$ to be signal
with \deltat\ distribution $\cal G$, defined in Eq.~\ref{eq:Phi},
is estimated from the $\mes$~fit (Fig.~\ref{fig:breco})
and the \mes\ value of the \Brec\ candidate.
Each event~{\it i} then samples a PDF that
includes signal, background, and outlier components:\\[-6mm] 
\begin{eqnarray}
\label{eq:totalpdf}
\lefteqn{ {\cal F} (\deltat_i ,\sigma_i ,p_i^{sig} | \tau ; \hat{a}, \hat{b},
                  f_{out}^{sig},f_{out}^{bkg} ) = } \\
& & p_i^{sig} \cdot [(1-f_{out}^{sig}) \cdot  
                      {\cal G} (\deltat_i , \sigma_i |\tau , \hat{a}) 
                  + f_{out}^{sig} \cdot {\cal O}  (\deltat_i)  ] +
                  \nonumber \\
& &(1 - p_i^{sig}) \cdot [(1-f_{out}^{bkg}) \cdot 
                      {\cal B} (\deltat_i | \hat{b})
		  + f_{out}^{bkg} \cdot {\cal O} (\deltat_i)  ].
                   \nonumber 
\end{eqnarray}
The background \deltat\ distribution, $\cal B$, for each \B\ species
is modeled by the sum of a prompt component and a lifetime
component convoluted with a resolution function of the form given in
Eq.~\ref{eq:resol}, but with a separate set of parameters.
The fraction of non-prompt background, its effective lifetime 
and the background resolution parameters are determined
separately for charged and neutral \B\ mesons.
Signal and background outlier events have an assumed \deltat\
behavior ${\cal O}$ given by a Gaussian distribution with zero mean and a
fixed 10\ps\ width. The fractions of outliers in
signal and background are determined separately in the lifetime fit.

Since the same resolution function is used for neutral and
charged \B\ mesons, the fitting procedure maximizes
the log-likelihood function $\ln \cal L$ formed from the sum 
of two terms, one for each \B\ meson species, with common parameters 
$\hat{a}$ for $\cal R$:
\vspace*{-1.5mm}
\begin{eqnarray}
\small
\label{eq:loglik}
\lefteqn{\ln \cal L \it =    
\sum_{i+} \ln[{\cal F}  (\deltat_{i+} ,\sigma_{i+} ,p_{i+}^{sig} | 
                    \tau_{\Bu } ; \hat{a}, \hat{b}_+,
                    f_{out}^{sig,+},f_{out}^{bkg,+} )]  } \nonumber \\
& + & \sum_{i0} \ln[{\cal F}  (\deltat_{i0} ,\sigma_{i0} ,p_{i0}^{sig} | 
                    \tau_{\Bz } ; \hat{a}, \hat{b}_0,
                    f_{out}^{sig,0},f_{out}^{bkg,0} )].\;\;\;\;\;
\end{eqnarray}
The likelihood fit involves 19 free parameters. The parameter
$\tau_{\Bu }$ is replaced with
$\tau_{\Bu } = r \cdot \tau_{\Bz }$ to estimate the statistical error
on the lifetime ratio~$r$. The lifetime values were kept hidden 
until the event selection and the \deltat~reconstruction method,
as well as the fitting procedure, were finalized and the systematic
errors were determined.

\begin{figure}[t]
\includegraphics[width=0.855\linewidth,height=1.04952\linewidth]{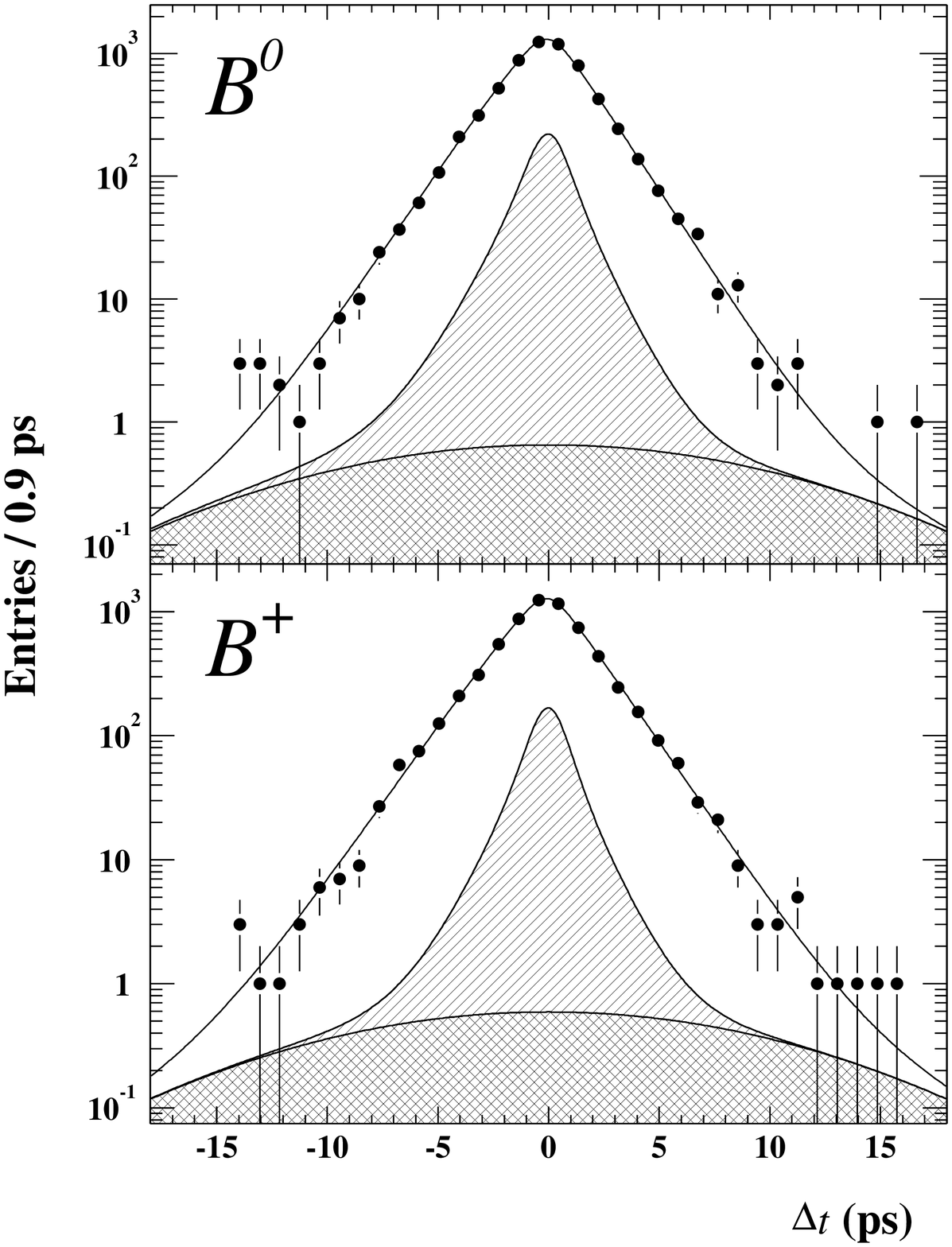}
\caption{\deltat\ distribution for the \Bz\ (top) and \Bu\ (bottom) 
events within $2.5 \sigma$ of the \B mass in \mes. The results of the fit are 
superimposed on the data. The single-hatched areas are the background
components $\cal B$ and the cross-hatched areas represent the outlier
contributions. The probability of obtaining a lower likelihood,
evaluated with a Monte Carlo technique, is 7.3\%.
\label{fig:blifefit}
}
\end{figure}

The fit results, after small
corrections discussed below, are
$\tau_{\Bz} = 1.546 \pm 0.032\ps$, 
$\tau_{\Bu} = 1.673 \pm 0.032\ps$ and
$\tau_{\Bu} / \tau_{\Bz} = 1.082 \pm 0.026$, where the
errors are statistical only.  
The resolution parameters
$\hat{a}$ ($h=0.69\pm0.07$, $s=1.21\pm0.07$ and $\kappa=1.04\pm0.24$)
are consistent with those found in a Monte Carlo simulation that 
includes detector alignment effects. The fitted outlier fractions 
in the \Bu\ and \Bz\ signals are both $0.2_{-0.2}^{+0.3}$\%. 
Figure~\ref{fig:blifefit} shows the results of the fit superimposed on the
observed \deltat~distributions for \Bz\ and \Bu\ events within 
2.5~standard deviations of the \B\ mass in \mes.

Table~\ref{tab:systematics} summarizes the systematic uncertainties on the
lifetime results.
\begin{table}[t!]
\caption{
\label{tab:systematics}
Summary of the systematic uncertainties.
}
\begin{tabular}{|l|c|c|c|}
\hline
Effect & $\delta(\tau_{\Bz})$ & $\delta(\tau_{\Bu})$ &
$\delta(\tau_{\Bu}/\tau_{\Bz})$  \\
&(ps) &(ps) & \\
\hline
\hline
MC statistics &  0.009   &  0.007 & 0.006 \\
$\cal R$ parametrization  &  0.008   &  0.004& 0.003 \\
same $\cal R$ for \Bz\ and \Bu\   &  0.004   &  0.005 & 0.006\\
Beam spot, $p_{\Brec}$ &  0.002   &  0.002 & cancels \\
\deltat\ outliers &  0.011   &  0.011& 0.005 \\
SVT alignment &  0.008   &  0.008& cancels \\
$z$ scale &  0.008   &  0.008 & cancels\\
\deltaz\ to \deltat\ conversion &  0.006   &  0.006 & cancels\\
Signal probability &  0.003   &  0.003 & 0.003 \\
Background modeling &  0.005   &  0.011& 0.005 \\
\hline
\hline
Total in quadrature  &  0.022   &  0.023& 0.012 \\
\hline
\end{tabular}
\end{table}
The full analysis chain, including event reconstruction and selection,
has been tested with Monte Carlo simulation. The statistical precision 
on the consistency between the generated and fitted lifetimes is assigned as
a systematic error.
The resolution parameters~$\hat{a}$
are determined from the data by the fit, contributing $\pm 0.017$\ps 
in quadrature to the statistical error of the individual lifetime results.
Thus, a large part of the \deltat~resolution uncertainty is included in
the statistical error. Residual systematic uncertainties are attributed to limited
flexibility of the resolution model. These contributions have been estimated
by comparing results with different
parametrizations.
We correct our measurements for the small positive (negative) 
bias on the \Bz (\Bu) lifetime due to differences in 
the \deltat\ resolution functions for \Bz~and \Bu~mesons 
arising from their decays to a different admixture of \Dm\ and \Dzb
mesons and estimated with a high-statistics Monte Carlo sample. 
The size of the correction is assigned
as a systematic error.
A small systematic error results from
uncertainties on the beam spot position
and vertical size, and the \Brec\ momentum vector, which are
used to constrain the \Bopp~vertex. 
To~estimate the systematic error due to the assumptions on the shape
of the \deltat\ outlier PDF, we first verified that the fitted
lifetime results are stable when distributions wider than 10\ps\  or 
even flat are used in the fit. To investigate narrower shapes 
which are more signal-like,
thousands of experiments with sets of fixed values for the
outlier width and mean were simulated and subjected to the nominal
lifetime fit. The largest observed bias is taken as 
systematic uncertainty. 
Additional systematic uncertainties are due to the SVT~alignment. 
The $z$~length scale was determined to better than 0.5\% from secondary 
interactions in a beam pipe section of known length.
Approximations in the calculation of \deltat\ from
\deltaz\ and the uncertainty on the boost lead to small systematic 
errors.
The errors on the \mes\ fit parameters are used to determine the
uncertainty on $p^{sig}$ and the corresponding systematic error.
The main systematic uncertainties related to backgrounds arise from
changes in the background composition as a function of \mes. 
An additional contribution arises from a 1-2\% \Bz~contamination of
the \Bu\ signal sample and vice versa. We use Monte Carlo simulation 
to correct for these background effects and assign the sum in
quadrature of the corrections as systematic uncertainty.

In summary, the \Bz and \Bu meson lifetimes and their ratio have
been determined to be:
\begin{eqnarray}
\small
 \tau_{\Bz} &=& 1.546 \pm 0.032\mbox{ (stat)} \pm 0.022\mbox{ (syst)} \mbox{ ps,}        \nonumber \\
 \tau_{\Bu} &=& 1.673 \pm 0.032\mbox{ (stat)} \pm 0.023\mbox{ (syst)} \mbox{ ps, and}    \nonumber \\
 \tau_{\Bu}/\tau_{\Bz} &=& 1.082 \pm 0.026\mbox{ (stat)} \pm 0.012\mbox{ (syst)}. \nonumber 
\normalsize
\end{eqnarray}
These are the most precise measurements to date, 
and they are consistent with the current world averages.

We are grateful for the 
extraordinary contributions of our \pep2\ colleagues in
achieving the excellent luminosity and machine conditions
that have made this work possible.
The collaborating institutions wish to thank 
SLAC for its support and the kind hospitality extended to them. 
This work is supported by
DOE
and NSF (USA),
NSERC (Canada),
IHEP (China),
CEA and
CNRS-IN2P3
(France),
BMBF
(Germany),
INFN (Italy),
NFR (Norway),
MIST (Russia), and
PPARC (United Kingdom). 
Individuals have received support from the Swiss NSF, 
A.~P.~Sloan Foundation, 
Research Corporation,
and Alexander von Humboldt Foundation.

\end{document}